\documentclass[12pt,a4paper]{article}
\usepackage{epsfig}
\begin{document}
\title{The neutral top-pion and the lepton flavor violating \\
$Z$ decays $Z\rightarrow l_{i}l_{j}$}
\author{Chongxing Yue, Wei Wang, Feng Zhang\\
{\small Department of Physics, Liaoning  Normal University, Dalian
116029. P.R.China}
\thanks{E-mail:cxyue@lnnu.edu.cn}
}
\date{\today}
\maketitle

\begin{abstract}
\hspace{5mm}Taking into account the constraints of the present
experimental limit of  the process $\mu\rightarrow e\gamma$ on the
free parameters of topcolor-assisted technicolor(TC2) models, we
calculate the contributions of the neutral top-pion to the lepton
flavor violating(LFV) $Z$ decays $Z\rightarrow l_{i}l_{j}$. Our
results show that the value of the branching ratio
$Br(Z\rightarrow \tau l)$ is larger than that of $Br(Z\rightarrow
\mu e)$ in all of the parameter space and there is
$Br(Z\rightarrow \tau\mu)\approx Br(Z\rightarrow \tau e)\simeq
2\times 10^{-14} $, which is far from the reach of present or
future experiments.
\end {abstract}

\vspace{2.0cm} \noindent
 {\bf PACS number(s)}:12.60Nz, 14.80.Mz, 12.14.65.Ha

\newpage

It is well known that the individual lepton numbers $L_{e}$,
$L_{\mu}$ and $L_{\tau}$ are automatically conserved and the
tree-level lepton flavor violating(LFV) processes are absent in
the standard model(SM). However, the solar neutrino experiments[1]
and the atmospheric neutrino experiments[2] confirmed by reactor
and accelerator experiments[3] provide very strong evidence for
mixing and oscillation of the flavor neutrinos, which presently
provide the only direct observation of physics that can not be
accommodated within the SM and imply that the separated lepton
numbers are not conserved. Thus, the SM requires some modification
to account for the pattern of neutrino mixing, in which the LFV
processes are allowed. The observation of these LFV processes
would be a clear signature of new physics beyond the SM at the
present or future experiments. This fact has made one to be of
interest in the LFV processes. Among them, the LFV $Z$ decays,
such as $Z\rightarrow l_{i}l_{j}$($l_{i}= e, \mu$ or $\tau$), are
interest subjects. Furthermore, the Giga $Z$ option of the TESLA
linear collider project will work at the resonance and increase
the production rate of $Z$ boson[4]. This forces one to study the
LFV $Z$ decays precisely.

To completely avoid the problems arising from the elementary Higgs
field in the SM, various kinds of dynamical electroweak symmetry
breaking(EWSB) models have been proposed, and among which the
topcolor scenario is attractive because it explains the large top
quark mass and provides possible dynamics of EWSB.
Topcolor-assisted technicolor(TC2) models[5], flavor-universal TC2
models[6], top see-saw models[7], and top flavor see-saw models[8]
are four of such examples. The presence of the physical top-pions
in the low-energy spectrum is an inevitable feature of these kinds
of models[9]. Studying the possible signatures of the top-pions at
present and future high- or low-energy colliders can help the
collider experiments to search for top-pions, test topcolor
scenario and further to probe EWSB mechanism.

The branching ratios of the LFV processes $Z\rightarrow
l_{i}l_{j}$ are extremely small even in the SM with massive
neutrinos: $Br(Z\rightarrow l_{i}l_{j})\leq 10^{-54}$[10], which
are far from the reach of present or future experiments. The
current experimental limits obtained at LEPI are[11]:
\begin{eqnarray}
Br(Z\rightarrow \mu e)&<&1.7\times 10^{-6},\nonumber\\
Br(Z\rightarrow \tau e)&<&9.8\times 10^{-6},\\
Br(Z\rightarrow \tau \mu)&<&1.2\times 10^{-5},\nonumber
\end{eqnarray}
and with the improved sensitivity at TESLA, these limits could be
pulled down to [12]:
\begin{eqnarray}
Br(Z\rightarrow \mu e)&<&2\times 10^{-9},\nonumber\\
Br(Z\rightarrow \tau e)&<&f\times 1.5\times 10^{-8},\\
Br(Z\rightarrow \tau \mu)&<&f\times 2.2\times 10^{-8},\nonumber
\end{eqnarray}
with $f=0.2\sim 1.0$.

Recently, there are many studies on the LFV $Z$ decays
$Z\rightarrow l_{i}l_{j}$ in various models[13]. For example,
these processes are investigated in a model independent way[14],
supersymmetric models[15], the general two Higgs doublet
model[16], the Zee model[17], and theories with a heavy boson
$Z'$[18]. In Ref.[19], we study the contributions of the
non-universal gauge boson $Z'$ to the LFV $Z$ decays $Z\rightarrow
l_{i}l_{j}$. We find that the branching ratios $Br(Z\rightarrow
\tau e)$ and $Br(Z\rightarrow \mu e)$ can approach the
experimental upper limits in a sizable of the parameter space of
the flavor-universal TC2 models. The aim of this paper is to
consider the contributions of the neutral top-pion $\pi_{t}^{0}$
to the LFV $Z$ decays $Z\rightarrow l_{i}l_{j}$ in the context of
TC2 models, and see whether $\pi_{t}^{0}$ can give significant
contributions on these processes.

For TC2 models[5], TC interactions play a main role in breaking
the electroweak symmetry. Topcolor interactions make small
contributions to EWSB and give rise to the main part of the top
quark mass, $(1-\varepsilon)m_{t}$, with the parameter
$\varepsilon<< 1$. Thus, there is the following relation:
\begin{equation}
\nu_{\pi}^{2}+F_{t}^{2}=\nu^{2}_{W},
\end{equation}
where $\nu_{\pi}$ represents the contributions of TC interactions
to EWSB, $\nu_{W}=\nu/\sqrt{2}\approx 174GeV$, and $F_{t}=50GeV$
is the physical top-pion decay constant. This means that the
masses of the gauge bosons $W$ and $Z$ are given by absorbing the
linear combination the top-pions and technipions. The orthogonal
combination of the top-pions and technipions remains unabsorbed
and physical. However, the absorbed Goldstone linear combination
is mostly the technipions while the orthogonal combination is
mostly the top-pions, which are usually called physical
top-pions($\pi_{t}^{\pm}, \pi_{t}^{0}$). The flavor diagonal(FD)
couplings of the neutral top-pion $\pi_{t}^{0}$ to leptons can be
written as[9,20]:
\begin{equation}
\frac{m_{l}}{\nu}\overline{l}\gamma^{5}l\pi_{t}^{0},
\end{equation}
where $l=\tau, \mu$, and $ e$.

\begin{figure}[htb]
\vspace{-2cm}
\begin{center}
\epsfig{file=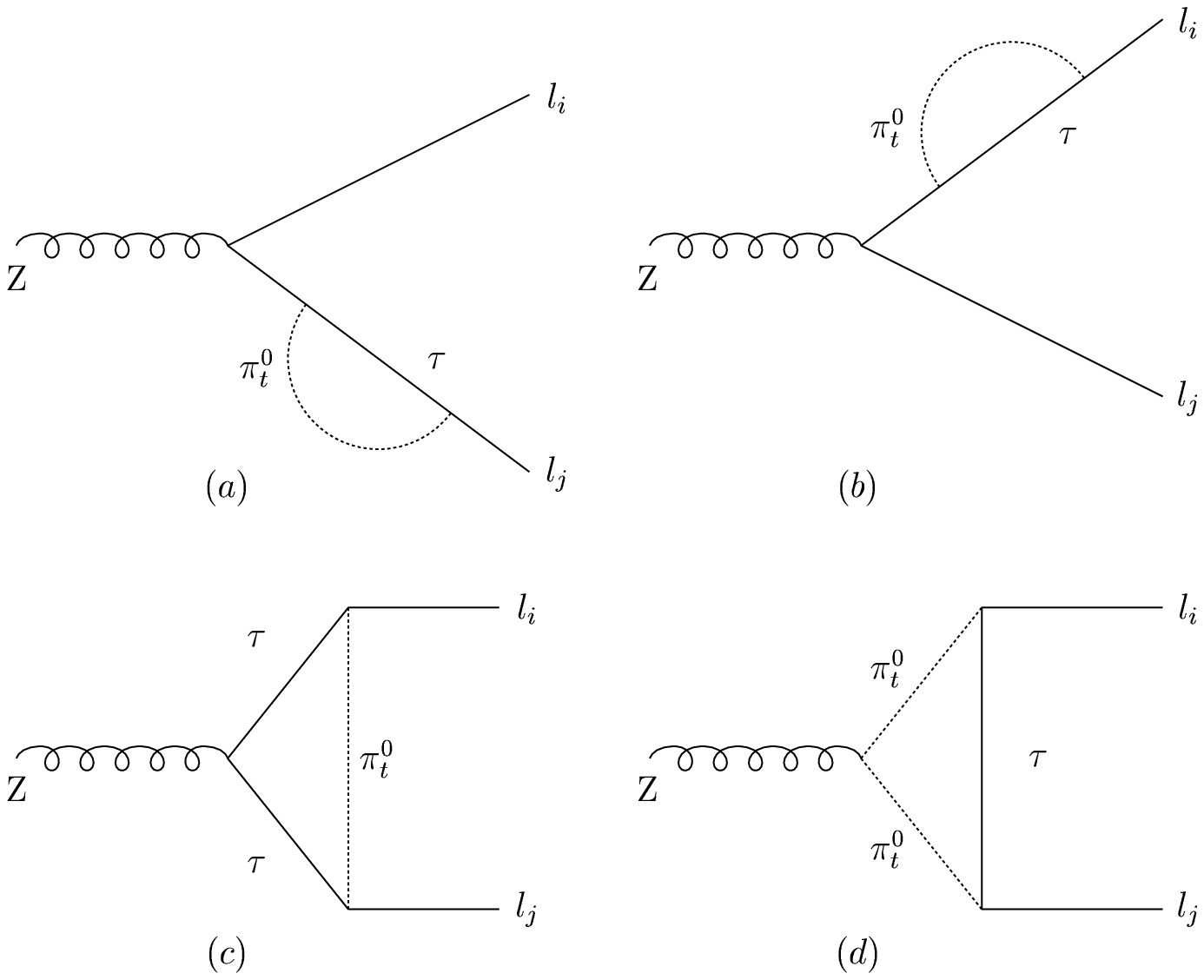,width=520pt,height=700pt} \vspace{-13.0cm}
\hspace{0.5cm} \caption{Feynman diagrams contribute to the $LFV$
processes $Z\rightarrow l_{i}l_{j}$ due to the neutral
\hspace*{1.7cm} top-pion $\pi^{0}_{t}$ exchange.} \label{ee}
\end{center}
\end{figure}

For TC2 models, the underlying interactions, topcolor
interactions, are non-universal and therefore do not posses GIM
mechanism. The non-universal gauge interactions result in the
flavor changing(FC) coupling vertices when one writes the
interactions in the mass eigenbasis. Thus, the top-pions can
induce the new FC scalar coupling vertices[21]. The FC couplings
of $\pi_{t}^{0}$ to leptons can be written as:
\begin{equation}
\frac{m_{\tau}}{\nu} k_{\tau i}
\overline{\tau}\gamma^{5}l_{i}\pi_{t}^{0},
\end{equation}
where $l_{i}(i=1,2)$ is the first(second) lepton $e(\mu)$,
$k_{\tau i}$ is the flavor mixing factor, which is the free
parameter. Certainly, there is also the FC scalar coupling
$\pi_{t}^{0} \mu e$. However, the topcolor interactions only
contact with the third generation fermions. The flavor mixing
between the first and second generation fermions is very small,
which can be ignored.

From above discussions, we can see that the $\pi_{t}^{0}$ can
indeed induce the LFV $Z$ decays $Z\rightarrow l_{i}l_{j}$ via the
FC scalar couplings $\pi_{t}^{0}\tau l_{i}$. The relevant Feynman
diagrams are depicted in Fig.1. The internal fermion line may be
the leptons $\tau,\mu$ or $e$. However, the internal fermion
propagator provides a term proportional to $m_{l}^{2}$ in the
numerator, which is not cancelled by the $m_{l}^{2}$ in the
denominator since the heavy $\pi_{t}^{0}$ mass $m_{\pi_{t}}$
dominates the denominator. Thus, we only take the internal fermion
line as the $\tau$ lepton line.

Using Eq.(4), Eq.(5) and other relevant Feynman rules, the
effective vertex $Z\overline{\mu}e$ contributed by $\pi_{t}^{0}$
exchange can be written as:
\begin{equation}
\Lambda_{Z\overline{\mu}e}=\frac{i e k_{\tau\mu}k_{\tau e}
m_{\tau}^{2}}{16\pi^{2}\nu^{2}}[\gamma^{\mu}(F_{1}+F_{2}\gamma^{5})+
P_{\mu}^{\mu}(F_{3}+F_{4}\gamma^{5})
+P_{e}^{\mu}(F_{5}+F_{6}\gamma^{5})],
\end{equation}
where $S_{W}=\sin \theta_{W}$, $\theta_{W}$ is the Weinberg angle.
The form factor $F_{i}$ can be written as:
\begin{eqnarray}
F_{1}&=&\frac{4S_{W}^{2}-1}{4S_{W}C_{W}}[-\frac{m_{e}}{m_{\mu}}B'_{1}-
\frac{m_{\tau}}{m_{\mu}}B'_{0}
-B''_{1}+\frac{m_{\tau}-m_{\mu}}{m_{\mu}}B''_{0}-B'''_{0}-m_{\pi_{t}}^{2}C'_{0}\nonumber\\
&&-m_{\tau}(m_{\tau}-m_{\mu})C'_{0}+2C'_{24}-m_{e}(m_{\tau}-m_{\mu})(C'_{11}-C'_{12})
+m_{\tau}m_{\mu}C'_{12}\nonumber\\
&&-m_{\tau}m_{e}(C'_{11}-C'_{12})+m_{\mu}(m_{\tau}-m_{\mu})C'_{12}],\nonumber\\
F_{2}&=&\frac{1}{4S_{W}C_{W}}[-\frac{m_{e}}{m_{\mu}}B'_{1}-\frac{m_{\tau}}{m_{\mu}}B'_{0}
-B''_{1}+\frac{m_{\tau}-m_{\mu}}{m_{\mu}}B''_{0}+B'''_{0}+m_{\pi_{t}}^{2}C'_{0}\nonumber\\
&&-m_{\tau}(m_{\tau}-m_{\mu})C'_{0}-2C'_{24}-m_{e}(m_{\tau}-m_{\mu})(C'_{11}-C'_{12})
+m_{\tau}m_{\mu}C'_{12}\nonumber\\
&&+m_{\tau}m_{e}(C'_{11}-C'_{12})-m_{\mu}(m_{\tau}-m_{l_{i}})C'_{12}-4C''_{24}],\nonumber\\
F_{3}&=&\frac{4S_{W}^{2}-1}{4S_{W}C_{W}}[-2m_{e}(C'_{22}-C'_{23})+2m_{\mu}C'_{22}
-2(m_{\tau}-m_{\mu})C'_{12}],\\
F_{4}&=&\frac{1}{4S_{W}C_{W}}[-2m_{e}(C'_{22}-C'_{23})-2m_{\mu}C'_{22}+2(m_{\tau}
-m_{\mu})C'_{12}+2m_{\tau}C''_{0}\nonumber\\
&&-2m_{e}(C''_{11}-C''_{12})+2m_{\mu}C''_{12}+4m_{\tau}C''_{12}+4m_{e}(C''_{22}
-C''_{23})+4m_{\mu}C''_{22}],\nonumber\\
F_{5}&=&\frac{4S_{W}^{2}-1}{4S_{W}C_{W}}[-2m_{e}(C'_{21}+C'_{22}-2C'_{23})
+2m_{\mu}(C'_{22}-C'_{23})-2m_{\tau}(C'_{11}-C'_{12})],\nonumber\\
F_{6}&=&\frac{1}{4S_{W}C_{W}}[-2m_{e}(C'_{21}+C'_{22}-2C'_{23})-2m_{\mu}(C'_{22}-C'_{23})
-2m_{\tau}(C'_{11}-C'_{12})\nonumber\\
&&-2m_{\tau}C''_{0}+2m_{e}(C''_{11}-C''_{12})-2m_{\mu}C''_{12}-4m_{\tau}(C''_{11}-C''_{12})\nonumber\\
&&+4m_{e}(C''_{21}+C''_{22}-2C''_{23})+4m_{\mu}(C''_{22}-C''_{23})].\nonumber
\end{eqnarray}
The two- and three-point Feynman integrals $B_{n},C_{0}$ and
$C_{ij}$ can be written as[22]:
\begin{eqnarray}
B'_{n}&=&B_{n}[-P_{e},m_{\pi_{t}},m_{\tau}],\hspace{1cm}B''_{n}=B_{n}
[-P_{\mu},m_{\pi_{t}},m_{\tau}],\nonumber\\
B'''_{n}&=&B_{n}[-P_{Z},m_{\tau},m_{\tau}],\hspace{1cm}C'_{0}=C_{0}
[P_{e},-P_{Z},m_{\pi_{t}},m_{\tau},m_{\tau}],\nonumber\\
C''_{0}&=&C_{0}[P_{e},-P_{Z},m_{\tau},m_{\pi_{t}},m_{\pi_{t}}],\\
C'_{ij}&=&C_{ij}[P_{e},-P_{Z},m_{\pi_{t}},m_{\tau},m_{\tau}],\nonumber\\
C''_{ij}&=&C_{ij}[P_{e},-P_{Z},m_{\tau},m_{\pi_{t}},m_{\pi_{t}}].\nonumber
\end{eqnarray}
Where $p_{\mu}$ and $p_{e}$ denote the momenta of the two final
state leptons $\mu$ and $e$, $p_{Z}$ denotes the momentum of the
gauge boson $Z$. For the case of calculating the contributions of
$\pi_{t}^{0}$ to the LFV $Z$ decays $Z\rightarrow l_{i}l_{j}$, the
process $Z\rightarrow \tau l$ (l=$\mu$ or $e$) is similarly to the
process $Z\rightarrow \mu e$. The differences between
$Z\rightarrow \mu e$ and $Z\rightarrow \tau l$ are the final state
particles and the FC scalar coupling forms. The neutral top-pion
$\pi_{t}^{0}$ generates the LFV process $Z\rightarrow \mu e$ via
the FC couplings $\pi_{t}^{0} \tau \mu$ and $\pi_{t}^{0} \tau e$,
but the LFV process $Z\rightarrow \tau e$ ($\tau \mu$) is induced
by the FC coupling $\pi_{t}^{0}\tau e$ ($\pi_{t}^{0}\tau \mu$).
Thus, the effective vertices $\Lambda_{Z\overline{\tau} e}$ and
$\Lambda_ {Z\overline{\tau} \mu}$ can be written as:
\begin{eqnarray}
\Lambda_{Z\overline{\tau}e}&=&\Lambda_{Z\overline{\mu}e}(m_{\mu}\rightarrow
m_{\tau}, k_{\tau\mu}k_{\tau e}\rightarrow k_{\tau e}),\\
\Lambda_{Z\overline{\tau}\mu}&=&\Lambda_{Z\overline{\tau}e}(m_{e}\rightarrow
m_{\mu}, k_{\tau e}\rightarrow k_{\tau \mu}).
\end{eqnarray}
In above equations, we have taken into account all the masses of
internal lepton $\tau$ and external leptons(anti-leptons).

In general, the decay width of the $LFV$ process $Z\rightarrow
l_{i}l_{j}$ can be written as:
\begin{equation}
\Gamma=\int
\frac{(2\pi)^{4}}{6m_{Z}}\delta^{4}(P_{Z}-q_{1}-q_{2})\frac{d^{3}q_{1}}{(2\pi)^{3}2E_{1}}
\frac{d^{3}q_{2}}{(2\pi)^{3}2E_{2}}|M|^{2}(P_{Z},q_{1},q_{2}),
\end{equation}
where $M$ is the amplitude of the process $Z\rightarrow
l_{i}l_{j}$. $q_{1}$ and $q_{2}$ denote the momenta of the two
final state leptons.

To obtain numerical results, we take the SM parameters as
$\alpha(m_{Z})=\frac{1}{128.8}$, $S_{W}^{2}=0.2315$,
$m_{\tau}=1.777GeV$, $m_{\mu}=0.105GeV$ and $m_{e}$=0 [11]. The
limits on the mass $m_{\pi_{t}}$ of the top-pion may be obtained
via studying its effects on the various experimental
observables[9]. It has been shown that $m_{\pi_{t}}$ is allowed to
be in the range of a few hundred $GeV$ depending on the models. As
numerical estimation, we take the top-pion mass $m_{\pi_{t}}$ as a
free parameter.
\begin{figure}[htb]
\vspace{0cm}
\begin{center}
\epsfig{file=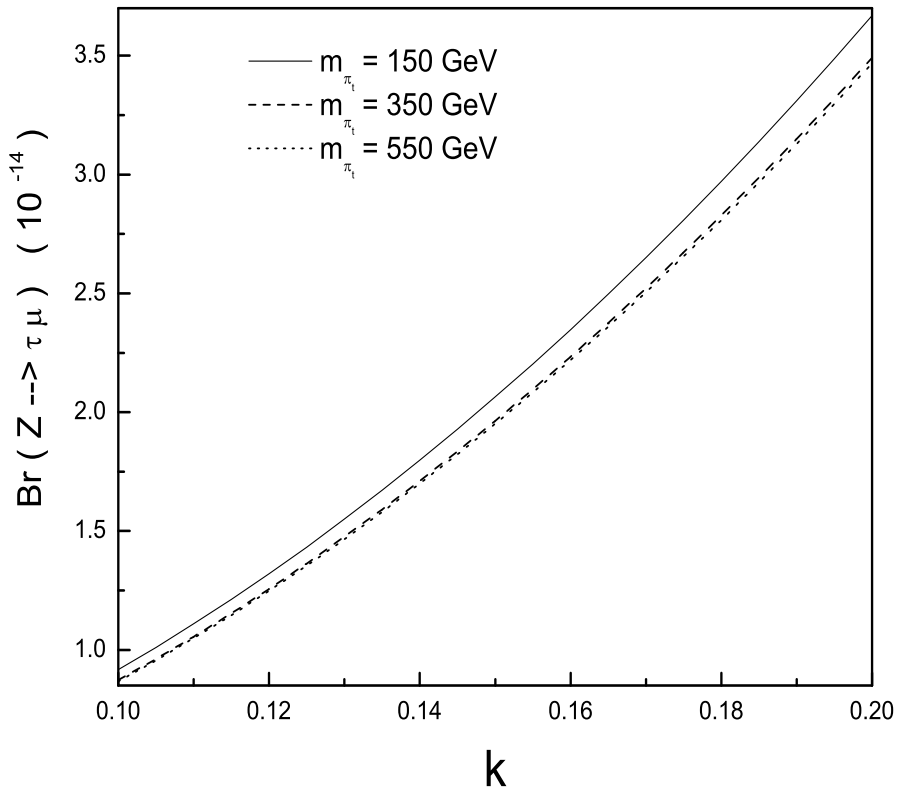,width=380pt,height=350pt} \vspace{-1cm}
\hspace{0.5cm} \caption{The branching ratios of the LFV process
$Z\rightarrow \tau \mu$ as a function of the flavor
\hspace*{1.8cm} mixing factor $k$ for three values of the top-pion
mass $m_{\pi_{t}}$.} \label{ee}
\end{center}
\end{figure}

For TC2 models, the topcolor interactions only contact with the
third generation. The new particles, such as extra gauge boson
$Z'$ and top-pions $\pi_{t}^{0,\pm}$, treat the fermions in the
third generation differently from those in the first and second
generation and treat the fermions in the first generation same as
those in the second generation. So, we can assume that the flavor
mixing factor $k_{\tau \mu}$ is equal to the flavor mixing factor
$k_{\tau e}$: $k=k_{\tau \mu}=k_{\tau e}$. In this case, we have
$Br(Z\rightarrow \tau\mu)\approx Br(Z\rightarrow \tau e)$ for
$m_{\mu}\approx 0$ and $m_{e}\approx 0$.

\begin{figure}[htb]
\vspace{1cm}
\begin{center}
\epsfig{file=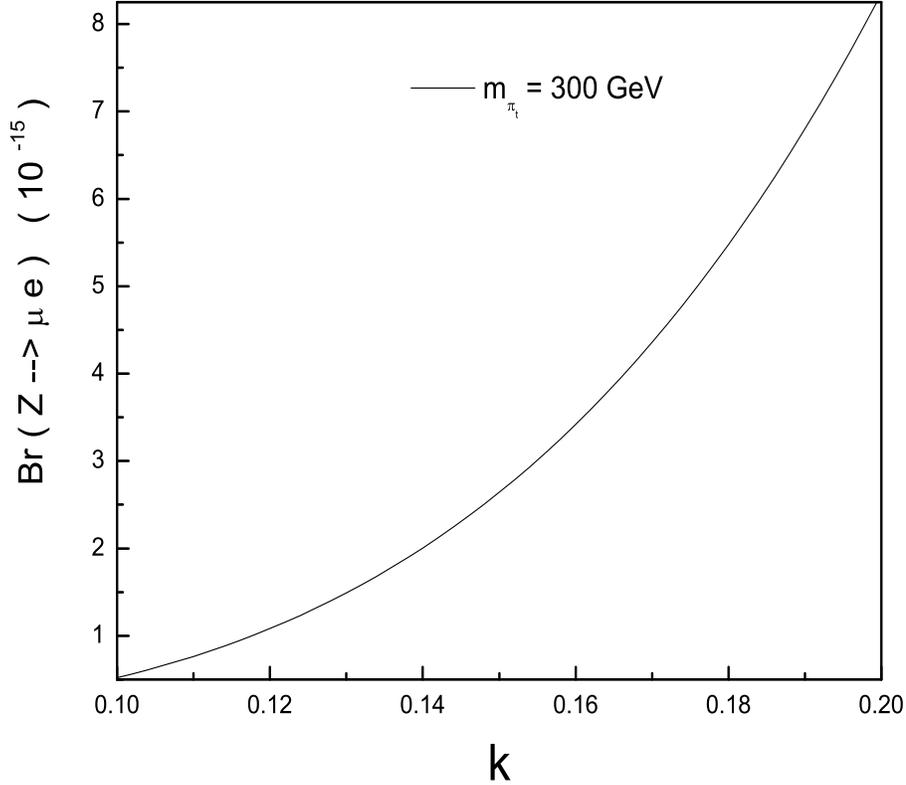,width=380pt,height=350pt} \vspace{-1cm}
\hspace{0.5cm} \caption{The branching ratio $Br(Z\rightarrow \mu e
)$ as a function of the flavor mixing factor $k$ for
\hspace*{1.8cm}$m_{\pi_{t}}=300GeV$.} \label{ee}
\end{center}
\end{figure}

The neutral top-pion $\pi_{t}^{0}$ can give significant
contributions to the LFV proscess $l_{i}\rightarrow l_{j}\gamma$
and $l_{i}\rightarrow l_{j}l_{k}l_{l}$[23]. The present
experimental bound on $\mu\rightarrow e \gamma$ gives severe
constraint on the mixing factor $k$. If we assume that the
$\pi_{t}^{0}$ mass $m_{\pi_{t}}$ is smaller than $400GeV$, then
there must be $k\leq 0.2$[23]. Taking into account the constraint
on the parameter $k $, the branching ratio $Br(Z\rightarrow \tau
l)(l=\mu$ or $e)$ is plotted in Fig.2 as a function of the
 parameter $k$ for three values of the top-pion mass $m_{\pi_{t}}$.
From Fig.2 one can see that the value of the branching
$Br(Z\rightarrow \tau l)$ increases as $k$ increasing and is
insensitive to $m_{\pi_{t}}$. For $200GeV\leq m_{\pi_t}\leq
400GeV$ and $0.1\leq k\leq 0.2$, the branching ratio
$Br(Z\rightarrow \tau l)$ is in the range of $8.7\times
10^{-15}\sim 3.6\times 10^{-14}$. The branching ratio
$Br(Z\rightarrow \mu e)$ is shown in Fig.3 as a function of $k$
for $m_{\pi_{t}}=300GeV$. For $ k =0.2$, the value of the
branching ratio $Br(Z\rightarrow \mu e )$ can reach $8.3\times
10^{-15}$.

In general, topcolor scenario predicts the existence of the
non-universal gauge boson $Z'$ and the top-pions
$\pi_{t}^{\pm,0}$. These new particles can induce the FC
couplings, which have significant contributions to the LFV
processes. In Ref.[19], we have studied the contributions of the
non-universal gauge boson $Z'$ to the LFV $Z$ decays $Z\rightarrow
l_{i}l_{j}$ in the context of the flavor-universal TC2 models and
TC2 models. Considering the constraint of the $B\overline{B}$
mixing on the $Z'$ mass $M_{Z'}$, we find that, in most of the
parameter space of TC2 models, there are $Br(Z\rightarrow \tau
\mu)\approx Br(Z\rightarrow \tau e)<1 \times 10^{-11}$ and
$Br(Z\rightarrow \mu e)<1 \times 10^{-13}$. Thus, the
contributions of the TC2 models to the LFV $Z$ decays
$Z\rightarrow l_{i}l_{j}$ mainly come from the non-universal gauge
boson $Z'$. This is because the couplings of the neutral top-pion
$\pi_{t}^{0}$ to leptons are very small. They are proportional to
the factor $\frac{m_{l}}{\nu}$, in which $m_{l}$ is the lepton
mass and $\nu\approx$ 246 $GeV$.

 High energy $e^{+}e^{-}$ colliders can be used as $Z$ factory,
providing an opportunity to examine the decay properties of the
gauge boson $Z$ in detail. The improved experimental measurements
at present stimulate the studies of the $Z$ decays. For example,
with the Giga-$Z$ option of the TESLA linear collider project, one
may expect the production of about $10^{9} Z$ bosons at
resonance[4]. The huge rate allows one to study a number of
problems with unprecedented precision. Among them is the search
for the LFV $Z$ decays $Z\rightarrow l_{i}l_{j}$. Ref.[23] has
shown that the present experimental bound on the LFV process
$\mu\rightarrow e \gamma$ produce severe constraints on the free
parameters of TC2 models. In this paper, based on these
constraints, we calculate the contributions of the neutral
top-pion $\pi_{t}^{0}$ to the LFV $Z$ decays $Z\rightarrow
l_{i}l_{j}$ via the FC scalar couplings $\pi_{t}^{0} \tau l_{i}$.
We find that the branching ratio $Br(Z\rightarrow \mu e)$ is
smaller than the branching ratio $Br(Z\rightarrow \tau \mu)$. In
wide range of parameter space of TC2 models, there is
$Br(Z\rightarrow \tau e)\approx Br(Z\rightarrow \tau \mu) \sim
10^{-14}$, which is far from the reach of present or future
experiments.

\vspace{1.5cm} \noindent{\bf Acknowledgments}

This work was supported by the National Natural Science Foundation
of China (90203005) and the Natural Science Foundation of the
Liaoning Scientific Committee(20032101).

\end{document}